%% file: eprint.tex
\documentclass[12pt]{article}
\usepackage{graphicx}

\newcommand\pubnumber{SNSN-323-63}
\newcommand\pubdate{\today}

\def\roma{Dipartimento di Fisica and INFN\\
Universit\'a di Roma "Sapienza", Piazzale A. Moro 5, 00185 Roma, Italy}
\def\genova{INFN, Sezione di Genova\\ Via Dodecaneso 33, 16146 Genova, Italy}
\def\mexico{Instituto de Ciencias Nucleares, Universidad Nacional Aut\'onoma de M\'exico \\ 04510 M\'exico DF, M\'exico}
\def\support{\footnote{Work supported by Universit\'a di Roma "Sapienza", 
          Italy, and INFN, Italy.}}

\def\Title#1{\begin{center} {\Large #1 } \end{center}}
\def\Author#1{\begin{center}{ \sc #1} \end{center}}
\def\Address#1{\begin{center}{ \it #1} \end{center}}

\newcommand\pubblock{\rightline{\begin{tabular}{l} \pubnumber\\
         \pubdate  \end{tabular}}}
\newenvironment{Abstract}{\begin{quotation}  }{\end{quotation}}
\newenvironment{Presented}{\begin{quotation} \begin{center} 
             PRESENTED AT\end{center}\bigskip 
      \begin{center}\begin{large}}{\end{large}\end{center} \end{quotation}}

\input econfmacros.tex

\begin{document}
\begin{titlepage}
\pubblock

\vfill
\Title{Higher charmonia and bottomonia. Nature of the $X(3872) $}
\vfill
\Author{Jacopo Ferretti\support}
\Address{\roma}
\Author{Elena Santopinto}
\Address{\genova}
\Author{Hugo Garc\'ia-Tecocoatzi}
\Address{\genova \\ \mexico}
\vfill
\begin{Abstract}
In this contribution, we will discuss our unquenched quark model results for the spectra of $c \bar c$ and $b \bar b$ states. The spectra are calculated in the unquenched quark model formalism, where the physical mass of a meson is the sum of a bare mass contribution plus a self-energy correction, due to the coupling to the meson-meson continuum. We will also discuss the quark structure of the $X(3872)$. Its open-flavor and radiative decays are calculated in the quark model formalism. In our interpretation, the wave function of the $X(3872)$ is the sum of two contributions, a $c \bar c$ core plus higher Fock components, due to meson-meson continuum components. Our results for the spectra, open-flavor and radiative decays are compared to the existing experimental data.
\end{Abstract}
\vfill
\begin{Presented}
CIPANP 2015\\
Vail (CO), USA, May 19--24, 2015
\end{Presented}
\vfill
\end{titlepage}
\def\thefootnote{\fnsymbol{footnote}}
\setcounter{footnote}{0}

\section{An unquenched quark model for mesons}
In the unquenched quark model (UQM) for mesons \cite{bottomonium,charmonium,charmonium02,Ferretti:2013vua} (see also Refs. \cite{Tornqvist,Geiger-Isgur,Burns:2014zfa}), a meson's wave function,
\begin{eqnarray} 
	\label{eqn:Psi-A}
	\mid \psi_A \rangle &=& {\cal N} \left[ \mid A \rangle 
	+ \sum_{BC \ell J} \int d \vec{q} \, \mid BC \vec{q} \, \ell J \rangle \right.
	\nonumber\\
	&& \hspace{2cm} \left.  \frac{ \langle BC \vec{q} \, \ell J \mid T^{\dagger} \mid A \rangle } 
	{E_a - E_b - E_c} \right] ~, 
\end{eqnarray}
is the superposition of a zeroth order quark-antiquark configuration plus a sum over the possible higher Fock components, due to the creation of $q \bar q$ sea pairs. 
Above threshold, this coupling to the meson-meson continuum leads to open-flavor strong decays; below threshold, it leads to virtual $q \bar q - q \bar q$ components in the meson wave function and a shift of the physical mass of the meson with respect to its bare mass, namely a self-energy correction. 

In this contribution, we will mainly focus on the calculation of self-energy corrections, $\Sigma(E_a)$.
They can be computed as \cite{bottomonium,charmonium,Ferretti:2013vua}
\begin{equation}
	\label{eqn:self-a}
	\Sigma(E_a) = \sum_{BC} \int_0^{\infty} q^2 dq \mbox{ } \frac{\left| \left\langle BC \vec q  \, \ell J \right| T^\dag \left| A \right\rangle \right|^2}{E_a - E_{bc}}  \mbox{ },
\end{equation}
where $E_a$ is a meson's bare mass, which is related by its physical mass by
\begin{equation}
	\label{eqn:self-trascendental}
	M_a = E_a + \Sigma(E_a)  \mbox{ }.
\end{equation}
Of particular interest are also the continuum components [see Eq. (\ref{eqn:Psi-A})]; they are given by \cite{bottomonium,charmonium02}
\begin{equation}
	\label{eqn:continuum}
	P_a^{sea} = \sum_{BC} \int_0^{\infty} q^2 dq \mbox{ } \frac{\left| \left\langle BC \vec q  \, \ell J \right| T^\dag \left| A \right\rangle \right|^2}{(E_a - E_{bc})^2}  \mbox{ },
\end{equation}
where one has to sum over meson-meson channels $BC$, such as $D\bar D^*$, and so on. The probability to find the meson in its core (or $q \bar q$) component is then
\begin{equation}
	P_a^{core}  = 1- P_a^{sea} \mbox{ }.
\end{equation}	

\begin{figure*}
\begin{tabular}{cc}
\includegraphics[width=3.in]{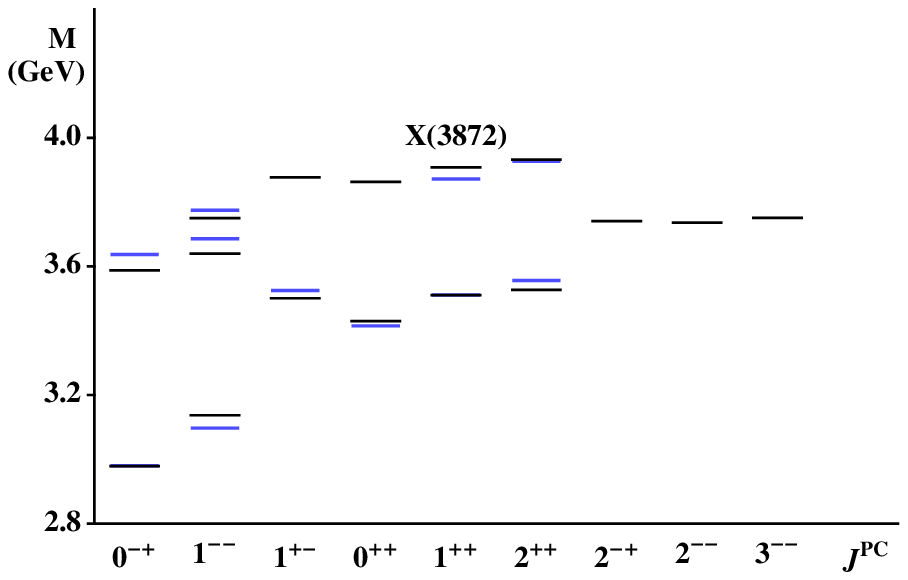} & \includegraphics[width=3.in]{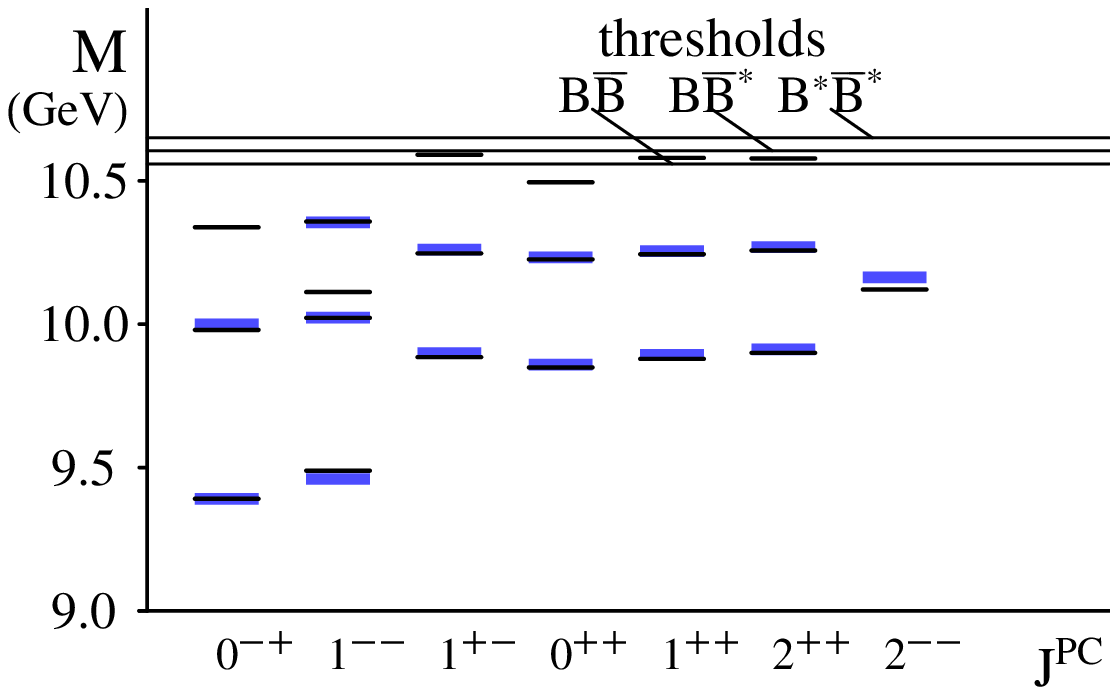}
\end{tabular}
\caption{Spectrum of $c \bar c$ (left panel) and $b \bar b$ mesons (right panel) in the UQM. The calculated masses are shown as black lines, the experimental data \cite{Nakamura:2010zzi} as blue boxes. Pictures from Refs. \cite{charmonium,Ferretti:2013vua}; APS copyright.}
\label{fig:spectra}
\end{figure*}

\section{Spectroscopy of higher charmonia and bottomonia}
In Refs. \cite{charmonium,Ferretti:2013vua}, the UQM was used to calculate the charmonium and bottomonium spectra with self-energy corrections. 
The physical masses of $c \bar c$ and $b \bar b$ mesons were calculated as in Eq. (\ref{eqn:self-trascendental}) and our theoretical results were fitted to the existing experimental data \cite{Nakamura:2010zzi}. The bare meson spectra were computed within Godfrey and Isgur's relativized quark model (QM) \cite{Godfrey:1985xj}. 

Our results are shown in Figs. \ref{fig:spectra}. It is worthwhile noting that the quality of the reproduction of the experimental data is good and, most important, that our predictions deviate from those of QMs in the case of states close to open-flavor decay thresholds, like the $X(3872)$ and $\chi_b(3P)$ mesons \cite{charmonium,charmonium02,Ferretti:2013vua}.
Our results were used to discuss the nature of the $X(3872)$ \cite{charmonium,charmonium02} and $\chi_b(3P)$ states \cite{charmonium02,Ferretti:2013vua}. In our UQM interpretation, these are quarkonium states plus higher Fock components due to the coupling to the meson-meson continuum. In this contribution, we will focus on the $X(3872)$.

\section{Quark structure of the $X(3872)$}
The $X(3872)$ is characterized by $J^{PC} = 1^{++}$ quantum numbers, a very narrow width (less than 1.2 MeV), and a mass $50-100$ MeV lower than QM predictions \cite{Nakamura:2010zzi,Choi:2003ue,Aaij:2013zoa}. 
There are several interpretations for this meson, including: 1) A "pure" $c \bar c$ state\footnote{For example, see the calculations of $c \bar c$ mesons' bare spectra of Ref. \cite{Godfrey:1985xj,Eichten:1978tg,Barnes:2005pb}.}; 2) A "pure" meson-meson molecule (for example, see Refs. \cite{Swanson:2003tb,Hanhart:2007yq,Aceti:2012cb}); 3) A tetraquark state \cite{Maiani:2004vq}; 4) A $c \bar c$ core plus higher Fock components, due to the coupling to the meson-meson continuum \cite{charmonium,charmonium02,Pennington:2007xr,Danilkin:2010cc,Karliner:2014lta,Cardoso:2014xda,Badalian:2015dha}. See also Ref. \cite{Padmanath:2015era}.

In the following, we will discuss our results of Refs. \cite{charmonium,charmonium02}.
There, we showed that the problem of the mass difference between QM predictions \cite{Godfrey:1985xj,Eichten:1978tg,Barnes:2005pb} and the experimental data for the $X(3872)$ can be solved thanks to a downward mass shift, due to self-energy (continuum coupling) effects. We will also show that the main decay modes of the $X(3872)$ are quite well reproduced in a "pure" $c \bar c$ description. The introduction of continuum coupling effects will require a renormalization of our results. This will be the subject of a subsequent paper \cite{subseq}.

\subsection{Radiative transitions}
Here, we discuss our QM results for the radiative transitions of the $X(3872)$ from Ref. \cite{charmonium02}. The $E1$ radiative transitions were calculated according to \cite{Eichten:1978tg}
\begin{equation}
	\label{eqn:radiative}
	\Gamma_{E1} = \frac{4}{3} C_{fi} \delta_{SS'} e_c^2 \alpha \left| \left\langle \psi_f \right| r \left| \psi_i \right\rangle \right|^2 E_\gamma^3 
	\frac{E_f^{(c \bar c)}}{M_i^{(c \bar c)}}  \mbox{ },
\end{equation}
where $e_c = \frac{2}{3}$ is the $c$-quark charge, $\alpha$ the fine structure constant, $E_f^{(c \bar c)}$ the total energy of the final $c \bar c$ state, $M_i^{(c \bar c)}$ the mass of the initial $c \bar c$ state, $E_\gamma$ the photon energy, $\left\langle \psi_f \right| r \left| \psi_i \right\rangle$ a radial matrix element, involving the initial and final mesons' radial wave functions, and the angular matrix element $C_{fi}$ is given by
\begin{equation}
	C_{fi} = \mbox{max} (L,L') (2J'+1) \left\{ \begin{array}{rcl} L' & J' & S \\ J & L & 1\end{array} \right\}^2 \mbox{ }.
\end{equation}
We calculated the matrix elements of Eqs. (\ref{eqn:radiative}) assuming, for the initial and final states, the wave functions of Godfrey and Isgur's relativized QM \cite{Godfrey:1985xj}. 

Finally, our results are reported in Table \ref{tab:radiative}. 

\begin{table}[t]
\begin{center}
\begin{tabular}{cccc}  
Transition & $\Gamma$(QM) [keV] & $\Gamma$(Molecule) [keV] & $\Gamma$(Exp.) [keV] \\ \hline
$X(3872) \rightarrow J/\Psi \gamma$                & 11  & 8 & $\approx 7$ \\
$X(3872) \rightarrow \Psi(2S) \gamma$            & 70  & 0.03 & $\approx 36$ \\
$X(3872) \rightarrow \Psi(3770) \gamma$         & 4.0 & 0 & $-$ \\
$X(3872) \rightarrow \Psi_2(1^3D_2) \gamma$ & 0.35 & 0 & $-$ \\ \hline
\end{tabular}
\caption{$E1$ radiative transitions of the $X(3872)$. Comparison between QM \cite{charmonium02} [see Eq. (\ref{eqn:radiative})], meson-meson molecular model results \cite{Swanson:2003tb}, and the experimental data \cite{Nakamura:2010zzi}.}
\label{tab:radiative}
\end{center}
\end{table}

\subsection{$D^0 \bar D^{0*}$ decay of the $X(3872)$}
In Ref. \cite{charmonium02}, we calculated the $D^0 \bar D^{0*}$ open-flavor decay of the $X(3872)$, also considering the possibility for the $\bar D^{0*}$ meson to decay then into $\bar D^0 \pi^0$. The width can be written as \cite{charmonium02,Segovia:2009zz}
\begin{equation}
	\begin{array}{l}
	\Gamma_{X(3872) \rightarrow D^0 (\bar D^0 \pi^0)_{\bar D^{0*}}} = \int_0^{q_{max}} dq \mbox{ } q^2 \mbox{ } \frac{2 \sum_{\ell J}
	| \langle D^0 \bar D^{0*} q \ell J | T^\dag | X(3872) \rangle |^2 \mbox{ } \Gamma(\bar D^{0*} \rightarrow \bar D^0 \pi^0;q)}{\left| M_a - E_b(q) - E_c(q) \right|^2 
	+ \frac{1}{4} \Gamma^2(\bar D^{0*})}  
	\end{array}  \mbox{ }.
\end{equation}
Here, $T^\dag$ is the $^3P_0$ model transition operator \cite{bottomonium,charmonium,charmonium02,Ferretti:2013vua,3P0,BandD}, $\Gamma(\bar D^{0*} \rightarrow \bar D^0 \pi^0;q)$ is the energy dependent decay amplitude of the unstable meson $\bar D^{0*}$, and $\Gamma^2(\bar D^{0*})$ is the total width of the $\bar D^{0*}$. Since the PDG \cite{Nakamura:2010zzi} only provides an upper limit for $\Gamma^2(\bar D^{0*}) < 2.1$ MeV, we took $\Gamma^2(\bar D^{0*})$ in the interval $0.25-2.1$ MeV. Our final result is
\begin{equation}
	\Gamma_{X(3872) \rightarrow D^0 (\bar D^0 \pi^0)_{\bar D^{0*}}} = 0.50 - 0.70  \mbox{ MeV },
\end{equation}
which is compatible with $\Gamma_{X(3872)} < 1.2$ MeV \cite{Nakamura:2010zzi}.

\section{Conclusion}
In this contribution, we discussed the results of an UQM calculation of the charmonium and bottomonium spectra with self-energy corrections \cite{bottomonium,charmonium,charmonium02,Ferretti:2013vua}. 
In the UQM \cite{bottomonium,charmonium,charmonium02,Ferretti:2013vua}, the effects of $q \bar q$ sea pairs are introduced explicitly into the QM via a QCD-inspired $^{3}P_0$ pair-creation mechanism \cite{3P0}. Below open-flavor decay thresholds, our UQM results are very similar to those of the relativized QM \cite{Godfrey:1985xj}; when meson energies approach meson-meson thresholds, QM and UQM results begin to split substantially.
In particular, this is the case of the $X(3872)$, whose mass is badly reproduced in QM calculations, while UQM results are closer to the experimental data.

In a second stage, we showed our QM results for the radiative transitions of the $X(3872)$ \cite{charmonium,charmonium02} and the $D^0 \bar D^{0*}$ decay. Our results are compatible with the present experimental data \cite{Nakamura:2010zzi}. A calculation of quarkonia radiative transitions and open-flavor decays in the UQM, including also higher Fock components in mesons wave functions, will be the subject of a subsequent paper.

\end{document}

%% file: econfmacros.tex
\def\beq{\begin{equation}}
\def\eeq#1{\label{#1}\end{equation}}
\def\eeqn{\end{equation}}

\def\beqa{\begin{eqnarray}}
\def\eeqa#1{\label{#1}\end{eqnarray}}
\def\eeqan{\end{eqnarray}}

\let\bar=\overbar

\def\Dslash{\not{\hbox{\kern-4pt $D$}}}
\def\dslash{\not{\hbox{\kern-2pt $\del$}}}

\def\msb{{\bar{\ssstyle M \kern -1pt S}}}




%% file: eprint.bbl
\begin{thebibliography}{99}


\bibitem{bottomonium}     
  J.~Ferretti, G.~Galat\'a, E.~Santopinto and A.~Vassallo,
  Phys.\ Rev.\ C \textbf{86}, 015204 (2012).

\bibitem{charmonium}
  J.~Ferretti, G.~Galat\'a and E.~Santopinto,
  Phys.\ Rev.\ C \textbf{ 88}, 015207 (2013).	
	
\bibitem{charmonium02}
  J.~Ferretti, G.~Galat\'a and E.~Santopinto,
  Phys.\ Rev.\ D \textbf{ 90}, 054010 (2014).

\bibitem{Ferretti:2013vua}   
  J.~Ferretti and E.~Santopinto,
  Phys.\ Rev.\ D \textbf{ 90}, 094022 (2014).
  
\bibitem{Tornqvist}
  N.~A.~T\"ornqvist and P.~Zenczykowski,
  Phys.\ Rev.\  D \textbf{ 29}, 2139 (1984);
  Z.\ Phys.\  C \textbf{ 30}, 83 (1986);
  P.~Zenczykowski,
  Annals Phys.\  \textbf{ 169}, 453 (1986).	
	
\bibitem{Geiger-Isgur}   
  P.~Geiger and N.~Isgur,
  Phys.\ Rev.\ Lett.\  \textbf{ 67}, 1066 (1991);
  Phys.\ Rev.\  D \textbf{ 44}, 799 (1991);
  \textbf{ 47}, 5050 (1993);
  \textbf{ 55}, 299 (1997).   
  
\bibitem{Burns:2014zfa} 
  T.~J.~Burns,
  Phys.\ Rev.\ D {\bf 90}, no. 3, 034009 (2014).  
  
\bibitem{Nakamura:2010zzi}
  K.~A.~Olive {\it et al.}  [Particle Data Group Collaboration],
  Chin.\ Phys.\ C {\bf 38}, 090001 (2014).
  
\bibitem{Godfrey:1985xj}
  S.~Godfrey and N.~Isgur,
  Phys.\ Rev.\ D \textbf{ 32}, 189 (1985).
	
\bibitem{Choi:2003ue}
  S.~K.~Choi {\it et al.}  [Belle Collaboration],
  Phys.\ Rev.\ Lett.\  {\bf 91}, 262001 (2003).
  
\bibitem{Aaij:2013zoa}
  R.~Aaij {\it et al.}  [LHCb Collaboration],
  Phys.\ Rev.\ Lett.\  {\bf 110}, 222001 (2013).  		
	
\bibitem{Eichten:1978tg}
  E.~Eichten, K.~Gottfried, T.~Kinoshita, K.~D.~Lane and T.~-M.~Yan,
  Phys.\ Rev.\ D \textbf{ 17}, 3090 (1978);
  \textbf{ 21}, 203 (1980).
	
\bibitem{Barnes:2005pb}
  T.~Barnes, S.~Godfrey and E.~S.~Swanson,
  Phys.\ Rev.\ D \textbf{ 72}, 054026 (2005).	
  
\bibitem{Swanson:2003tb}
  E.~S.~Swanson,
  Phys.\ Lett.\ B {\bf 588} (2004) 189;
  {\bf 598} (2004) 197.    
  
\bibitem{Hanhart:2007yq}
  C.~Hanhart, Y.~S.~Kalashnikova, A.~E.~Kudryavtsev and A.~V.~Nefediev,
  Phys.\ Rev.\ D {\bf 76} (2007) 034007.  
  
\bibitem{Aceti:2012cb} 
  F.~Aceti, R.~Molina and E.~Oset,
  Phys.\ Rev.\ D {\bf 86}, 113007 (2012).

\bibitem{Maiani:2004vq} 
  L.~Maiani, F.~Piccinini, A.~D.~Polosa and V.~Riquer,
  Phys.\ Rev.\ D {\bf 71}, 014028 (2005). 
			
\bibitem{Pennington:2007xr}
  M.~R.~Pennington and D.~J.~Wilson,
  Phys.\ Rev.\ D {\bf 76}, 077502 (2007).    

\bibitem{Danilkin:2010cc}
  I.~V.~Danilkin and Y.~A.~Simonov,
  Phys.\ Rev.\ Lett.\  {\bf 105}, 102002 (2010). 	
			
\bibitem{Karliner:2014lta} 
  M.~Karliner and J.~L.~Rosner,
  Phys.\ Rev.\ D {\bf 91}, no. 1, 014014 (2015).	
  
\bibitem{Cardoso:2014xda} 
  M.~Cardoso, G.~Rupp and E.~van Beveren,
  Eur.\ Phys.\ J.\ C {\bf 75}, no. 1, 26 (2015).  
  
\bibitem{Badalian:2015dha} 
  A.~M.~Badalian, Y.~A.~Simonov and B.~L.~G.~Bakker,
  Phys.\ Rev.\ D {\bf 91}, no. 5, 056001 (2015).  		
  
\bibitem{Padmanath:2015era} 
  M.~Padmanath, C.~B.~Lang and S.~Prelovsek,
  Phys.\ Rev.\ D {\bf 92}, no. 3, 034501 (2015).  
  
\bibitem{subseq}
   J.~Ferretti, H. Garc\'ia-Tecocoatzi and E.~Santopinto,	  
   in preparation.
    
\bibitem{Segovia:2009zz} 
  J.~Segovia, A.~M.~Yasser, D.~R.~Entem and F.~Fernandez,
  Phys.\ Rev.\ D {\bf 80}, 054017 (2009).	
  
\bibitem{3P0}
  L.~Micu,
  Nucl.\ Phys.\ B {\bf 10}, 521 (1969);
  A.~Le Yaouanc, L.~Oliver, O.~Pene and J.~-C.~Raynal,
  Phys.\ Rev.\ D {\bf 8}, 2223 (1973);
   {\bf 9}, 1415 (1974);
   R.~Kokoski and N.~Isgur,
  Phys.\ Rev.\ D {\bf 35}, 907 (1987);
  E.~S.~Ackleh, T.~Barnes and E.~S.~Swanson,
  Phys.\ Rev.\ D {\bf 54}, 6811 (1996).
  
\bibitem{BandD}
  J.~Ferretti, H. Garc\'ia-Tecocoatzi and E.~Santopinto,
  arXiv:1506.04415.       													
  
\end{thebibliography}
